\documentclass[aps,prd,twocolumn,showpacs,eqsecnum,nofootinbib]{revtex4}

\usepackage[dvips]{epsfig}
\usepackage{amssymb}
\usepackage{amsmath}
\usepackage{fancybox}

\begin{document}

\title{Stringent constraints on cosmological neutrino-antineutrino
asymmetries \\ from synchronized flavor transformation}
\author{Kevork N. Abazajian}
\author{John F. Beacom}
\author{Nicole F. Bell}
\affiliation{NASA/Fermilab Astrophysics Center, 
Fermi National Accelerator Laboratory, Batavia, Illinois 60510-0500\\
{\tt aba@fnal.gov, beacom@fnal.gov, nfb@fnal.gov}}
\date{March 25th, 2002}

\begin{abstract}
We assess a mechanism which can transform neutrino-antineutrino
asymmetries between flavors in the early universe, and confirm that
such transformation is unavoidable in the near bi-maximal framework
emerging for the neutrino mixing matrix.  We show that the process is
a standard Mikheyev-Smirnov-Wolfenstein flavor transformation dictated
by a synchronization of momentum states.  We also show that flavor
``equilibration'' is a special feature of maximal mixing, and
carefully examine new constraints placed on neutrino asymmetries.  In
particular, the big bang nucleosynthesis limit on electron neutrino
degeneracy $|\xi_e|\lesssim 0.04$ does not apply directly to all
flavors, yet confirmation of the large-mixing-angle solution to the
solar neutrino problem will eliminate the possibility of degenerate
big bang nucleosynthesis.
\end{abstract}
\pacs{14.60.Pq, 26.35.+c    \hspace{5cm} FERMILAB-Pub-02/056-A}

\maketitle


\section{Introduction}

As is well known, the observational successes of big bang
nucleosynthesis (BBN) are one of the pillars of standard
cosmology~\cite{schramm}.  If one assumes three standard neutrino
flavors, then the only free parameter is the baryon to photon ratio
$n_B/n_\gamma$.  The value $n_B/n_\gamma \simeq 5 \times 10^{-10}$
predicts light-element yields of $^2$H, $^4$He, and $^7$Li that are in
excellent agreement with observations.  As is often noted, this is
particularly remarkable because the absolute yields of these elements
span several orders of magnitude.  This consistency implies that the
post-BBN processing of the light elements is largely understood, and
that one does not require new aspects of particle physics beyond the
standard model (SM) that would materially affect BBN.

The basic consistency of our picture of the early universe is even more
impressive when one considers other recent cosmological measurements.
Observations of the acoustic peaks in the angular power spectrum of the cosmic
microwave background (CMB) give strong evidence that $\Omega_{total} = 1.04\pm
0.06$~\cite{cmb}, i.e., the universe is flat, as predicted by inflation.  Taken
together with measurements of clusters of galaxies and the high-redshift
type-Ia supernovae (SNIa) data, a mutually consistent picture~\cite{Turner}
with $\Omega_{matter} \simeq 0.3$ and $\Omega_{lambda} \simeq 0.7$ is obtained.
Additionally, the CMB data indicate that $\Omega_{baryon} \simeq 0.04$, in
excellent agreement with the BBN observations.  Recent data on the clustering
of galaxies also yield consistent values of $\Omega_{matter}$ and
$\Omega_{baryon}$~\cite{lss}.  This agreement of the combined data is all the
more impressive because baryonic matter is such a small fraction of the total
energy density of the universe, and because the BBN and CMB data reflect
measurements of $\Omega_{baryon}$ at very different epochs ($\sim 10^2$ s and
$\sim 10^{13}$ s after the big bang, respectively).

Nevertheless, from a particle-physics point of view, these
impressively measured quantities are all totally unexplained, both in
their values and their nature (e.g., though we know that the particle
dark matter is not part of the SM, we do not know what it is).  Just
as in accelerator-based particle physics, the underlying belief is
that more and more precise measurements will lead us to the necessary
clues on how to generalize the SM.  In particular, the
baryon-antibaryon asymmetry of $5 \times 10^{-10}$ remains a mystery,
and is certainly an important clue for understanding the universe at
temperatures at least as high as the electroweak scale.

Naturally, attention is also focused on the lepton-antilepton
asymmetry of the universe.  General considerations indicate that $B -
L$ may be conserved, so that the lepton asymmetry is $n_L/n_\gamma
\simeq 5 \times 10^{-10}$ as well.  However, there are certainly
viable models in which the lepton asymmetry can be much
larger~\cite{affleckdine}, and if confirmed, would be a very important
clue.  Given constraints on charge asymmetry, any large lepton
asymmetry would have to be hidden in the neutrino sector.  Though the
baryon asymmetry can generically be limited to be less than $10^{-8}$
simply to not overclose the universe, no similar constraints exist in
the lepton sector for light neutrinos.

Since neutrinos and antineutrinos should be in chemical equilibrium
until they decouple at a temperature $T \sim 2$ MeV, they may be
well described by Fermi-Dirac distributions with equal and opposite
chemical potentials:
\begin{equation}
f(p,\xi) = \frac{1}{1+\exp(p/T-\xi)}\,,
\end{equation}
where $p$ denotes the neutrino momentum, $T$ the temperature and $\xi$
is the chemical potential in units of $T$.  (There is a tiny
non-thermal perturbation that occurs at the epoch of $e^+ e^-$
annihilation at $T \simeq 0.3$ MeV, which we can ignore.)  The lepton
asymmetry $L_{\alpha}$ for a given flavor $\nu_{\alpha}$ is related to
the chemical potential by
\begin{equation}
L_{\alpha} = \frac{n_{\nu_\alpha} - n_{\bar{\nu}_\alpha}}{n_\gamma}
= \frac{\pi^2}{12\zeta(3)}\left(\xi_\alpha + 
\frac{\xi_\alpha^3}{\pi^2}\right)\,,
\label{leptxi}
\end{equation}
where $\zeta(3)\simeq 1.202$.  Even enormous values of $\xi_\alpha
\sim 1$ have been allowed observationally.  This is so distant from
the naive SM prediction that any measured nonzero value would be very
important.  Interest in searches for such large values of $\xi_\alpha$
is also driven by the fact that we evidently have much to learn about
the neutrino sector.  Previously, most attention was devoted to the
possibility of significant mixing of the SM active neutrinos with
light sterile neutrinos, which can generate large lepton numbers
$L\sim 1$~\cite{ftv}.

A general approach to setting limits on $\xi_\alpha$ arises because
for very large degeneracy, the effective number of neutrinos is
increased from the standard model prediction by
\begin{equation}
\Delta N_\nu = \frac{30}{7}\left(\frac{\xi}{\pi}\right)^2 +
\frac{15}{7}\left(\frac{\xi}{\pi}\right)^4.
\label{delnu}
\end{equation}
This increases the expansion rate of the universe, changing the CMB
results by magnifying the amplitude of the acoustic peaks.  For all
flavors, the bound $|\xi_\alpha| \lesssim 3$ has been obtained from
the CMB alone \cite{Hannestad:2000hc}.  Note that the sign of
$\xi_\alpha$ is unconstrained.  With future CMB data, these limits may
be reduced to $|\xi_\alpha|\lesssim 0.25$ or less~\cite{leptcmb}.  A
much stronger limit can be placed on $\xi_e$ with $|\xi_e| \lesssim
0.04$ because of its effect on setting the neutron to proton ratio
prior to BBN by altering beta equilibrium.\footnote{Beta equilibrium
is between the weak interactions $n+\nu_e\leftrightarrow p+e^-$ and
$p+\bar\nu_e\leftrightarrow n+e^+$. Positive $\xi_e$ increases the
$\nu_e$ abundance relative to $\bar{\nu}_e$, forcing equilibrium
towards lower $n/p$.}  If at the same time $\xi_{\mu,\tau}$ are
large, this effect can be partially undone by the increased expansion
rate, leading to the often-quoted bounds~\cite{hansen,kneller}
\begin{eqnarray}
\label{oldlimits}
-0.01 < \xi_e & < & 0.22\,, \\
|\xi_{\mu,\tau}| &<& 2.6\,,
\end{eqnarray}
where the upper limits are obtained only in tandem.

There are now three types of evidence for neutrino oscillations: solar
neutrino~\cite{solar} $\nu_e \rightarrow \nu_\mu,\nu_\tau$ with large (but not
maximal) mixing angle and $\delta m^2 \simeq 10^{-5}$ eV$^2$,
atmospheric~\cite{atm} neutrino $\nu_\mu,\bar\nu_\mu \rightarrow
\nu_\tau,\bar\nu_\tau$ with maximal mixing and $\delta m^2 \simeq 10^{-3}$
eV$^2$, and the Liquid Scintillator Neutrino Detector (LSND)~\cite{lsnd}
neutrino $\bar{\nu}_\mu \rightarrow \bar{\nu}_e$ with a very small angle and
$\delta m^2 \simeq 1$ eV$^2$.  It is not possible to accomodate these three
signals with only three neutrinos, as there are only two independent
mass-squared differences.  A possible fourth (sterile) neutrino can be invoked
to create a new $\delta m^2$, but now that the solar and atmospheric neutrino
data indicate the appearance of active neutrino flavors, there is a problem of
where to incorporate the required mixing with the sterile neutrino.  While
four-neutrino models may still work, it is only with difficulty, both in
fitting the oscillation data (see, e.g., Ref.~\cite{fournu}) and through the
effects on BBN (see, e.g., Ref.~\cite{dibarietal}).  The LSND signal will be
conclusively confirmed or refuted by the MiniBooNE experiment~\cite{miniboone}.
For simplicity, we consider just three active neutrinos, and neglect the LSND
result (of course, if it is confirmed, a major revision will be necessary).

In such a three-neutrino framework, Lunardini and Smirnov~\cite{smirnov}
suggested that the large mixing angles implied by the present data may transfer
any large asymmetry hidden in $\xi_{\mu,\tau}$ to $\xi_e$ well before the
beta-equilibrium freezeout at $T\simeq 1\rm\ MeV$ (see also Savage, Malaney and
Fuller~\cite{smf}).  Thus, the stringent BBN limit on $\xi_e$ might apply to
all three flavors, improving the bounds on $\xi_{\mu,\tau}$ by nearly two
orders of magnitude.

This proposal was recently studied in detail by Dolgov, Hansen, Pastor, Petcov,
Raffelt, and Semikoz (DHPPRS)~\cite{dhpprs}.  They found that close to
complete transformation of asymmetries $\xi_{\mu}$ and $\xi_{\tau}$ to $\xi_e$
was obtained.  This is an important result, as it excludes the possibility of
degenerate BBN~\cite{oritoesposito}, and is the strongest limit on the total
lepton number of the universe and is likely to remain so for the foreseeable
future.  In this article we examine the DHPPRS result, show it as the result of
a synchronized Mikheyev-Smirnov-Wolfenstein (MSW) transformation and establish
its robustness through physical and numerical insight into the dynamics.  We
assess how the results depend on the input parameters and consider more exotic
physical scenarios that might affect the results.


\section{Two-flavor density matrix equations and synchronized MSW}

In this section we consider a mixed neutrino statistical ensemble in the early
universe, with initial neutrino-antineutrino asymmetries which are not equal
among flavors. We show that this ensemble behaves as a synchronized system
following a single effective momentum state that undergoes an MSW
transformation given large mixing angles.  In describing neutrino flavor
evolution in dense environments such as the early universe, one must use a
density matrix description if the neutrino self-potential is large or if
decohering collisional processes are significant.  Where collisional processes
are not important, as is the case for some examples we shall consider here, the
evolution is coherent.  A useful parametrization of the density matrix
equations is the Bloch form \cite{stodolsky}.

In an environment such as the early universe, where the potential arising from
neutrino-neutrino forward scattering (the neutrino self-potential) is
important, active-active mixing is substantially different from active-sterile
mixing.  Forward scattering processes of the type $\nu_{\alpha}(p)\rightarrow
\nu_{\beta}(p)$ lead to refractive index terms which are off-diagonal in the
flavor basis $\{\alpha,\beta\}$~\cite{pantaleone}.  A useful and interesting
casting of the Bloch formalism for pure active neutrino mixing was done by
Pastor, Raffelt and Semikoz \cite{prs}, which allows an interpretation via
analogy with the precession of coupled magnetic dipoles.  The analysis of
Ref.~\cite{prs} considered the case of constant density, in the absence of a
background medium other than that provided by the neutrinos themselves. We
shall have need to extend this description to include a background medium of
charged leptons of a density that varies with time (or temperature).  One
particularly interesting feature first revealed clearly in Ref.~\cite{prs}, is
that the neutrino self-potential (that is, the potential due to
neutrino-neutrino forward scattering) does not, in general, suppress flavor
oscillations, as one might have naively expected by analogy with the potential
from, say, a background of charged leptons.  This is in stark contrast to the
case of active-sterile oscillations, where the effect of an asymmetry between
the active neutrinos and antineutrinos is always to suppress mixing angles.
Even for relatively small degeneracies, the asymmetry term dominates the
evolution and thus delays transformation of such asymmetry from the active to
the sterile flavor.

For active-active oscillations, no such simple mixing angle
suppression occurs as the neutrino asymmetry enters both the diagonal
and off-diagonal terms in the effective Hamiltonian. These terms have
the effect of synchronizing the ensemble,\footnote{Note that this
forward-scattering induced synchronization is unrelated to the
synchronization effect of Ref.~\cite{linearsync} which arises from
rapid flavor-blind collisions. It is remarkable that both collisions
and forward scattering lead to synchronization effects in the case of
active-active oscillations.}  resulting in collective behavior
resembling the evolution of a single momentum state in the absence of
the self-potential.  The mixing angle for this collective oscillation
is determined essentially by the background medium of thermal charged
leptons.  When the density of this background decreases with
temperature, the neutrinos evolve adiabatically from their initial
flavor into vacuum mass eigenstates.  For large-angle mixing, this
implies significant flavor transformation.


\subsection{Formulation}

We express the mixing between two neutrino mass and flavor eigenstates as
\begin{eqnarray}
\nu_e &=& \phantom{+}\cos \theta_0 \nu_1 +  \sin \theta_0 \nu_2\,,
\nonumber \\
\nu^\ast_\mu &=& -\sin \theta_0 \nu_1 +  \cos \theta_0 \nu_2,
\end{eqnarray}
where, in general, $\nu^\ast_\mu$ denotes some linear superposition of
$\nu_{\mu}$ and $\nu_{\tau}$, as we shall explain later.  We
parametrize the two-flavor neutrino density matrix in the form
\begin{eqnarray}
\rho(p) = \left( \begin{array}{cc}
\rho_{ee} & \rho_{e \mu} \\
 \rho_{\mu e} & \rho_{\mu \mu}
\end{array} \right)
=  \frac{1}{2} \left[ P_0(p)
+ \boldsymbol\sigma\cdot{\mathbf P}(p)\right],
\end{eqnarray}
and similarly for the antineutrinos, where we refer to ${\mathbf
P}(p)$ as the neutrino ``polarization'' vector.  These quantities are
most usefully normalized such that
\begin{equation}
\label{initcond}
{\mathbf P}(p)^{initial} = 
\frac{1}{n^{\rm eq}/T^3}
\left[f_e(p,\xi_e) - f_\mu(p,\xi_\mu)\right],
\end{equation}
where $n_0^{\rm eq}=\int d^3p/(2\pi)^3 f(p,0) = 3 \zeta(3) T^3/4 \pi^2$.

In the absence of collision terms, $P_0(p)$ and the magnitude of
${\mathbf P}(p)$ remain constant and the full evolution equations for
two mixed active flavors in the early universe are
\begin{eqnarray}
\label{veceqns}
\partial_t{\mathbf P}_p &=& +{\mathbf A}_p \times {\mathbf P}_p + \alpha
    ({\mathbf J}- {\mathbf{\overline J}})\times {\mathbf P}_p \,, \\
\nonumber \partial_t{\mathbf{\overline P}}_p &=& -{\mathbf{\overline
    A}}_p \times {\mathbf{\overline P}}_p + \alpha ({\mathbf J}-
    {\mathbf{\overline J}})\times {\mathbf{\overline P}}_p \,,
\end{eqnarray}
where ${\mathbf P}_p$ denotes the polarization vectors for the
neutrinos of momentum $p$ while ${\mathbf J}$ denotes the
corresponding quantity integrated over momentum such that
\begin{equation}
{\mathbf J} = \int \frac{d^3(p/T)}{(2\pi)^3} {\mathbf P}_p.
\end{equation}
Vectors with an overbar refer to the antineutrino quantities
throughout.  With the normalization we use here, the length of the
individual ${\mathbf P}_p$ vectors and that of ${\mathbf J}$ do not
redshift with temperature.  The coefficient of the second term is
$\alpha \equiv \sqrt{2} G_F n^{\rm eq}$.  Time $t$ and temperature $T$
may be interconverted via the expression $t\simeq 1.15{\rm\
s}\;(T/{\rm MeV})^{-2}$.  Decohering collisions (damping) of the
system at high temperatures forces the neutrinos into unmixed flavor
states.  We shall assume zero initial $\xi_e^i$ and a finite initial
$\xi_{\mu}^i$, taken to have a negative sign.\footnote{For the opposite
sign, one simply reverses the directions of the polarization vectors.}
Therefore, the initial alignment of the ${\mathbf P}_p$ are along the
$+z$ axis, and $\overline{\mathbf P}_p$ are along the $-z$ axis.

Equations~(\ref{veceqns}) are equivalent to the precession of magnetic dipoles
in two ``magnetic fields'': the momentum-dependent ${\mathbf A}_p$ and the
integrated neutrino self-potential $\alpha ({\mathbf J}- {\mathbf{\overline
J}})$.  The effects of ${\mathbf A}_p$, as we shall show, are straightforward,
but the neutrino self-potential makes the system non-linear by explicitly
coupling each momentum mode to the evolution of every other momentum mode.  An
intuitive description of the evolution in Eq.~(\ref{veceqns}) for a
constant-density system without matter effects was provided in Ref.~\cite{prs}.
The issue of the synchronization of the system has also been studied in
Ref.~\cite{Samuel}.

In general, the ``magnetic field'' vector ${\mathbf A}_p$ includes
contributions from vacuum mixing, a thermal potential from the
charged-lepton background, and a potential due to asymmetries between
the charged leptons
\begin{equation}
\label{magnetic}
{\mathbf A}_p ={\mathbf \Delta}_p + \left[V^T(p) +
V^B\right]\hat{\mathbf z}.
\end{equation}
Vacuum mixing is incorporated by
\begin{equation}
{\mathbf \Delta}_p = \frac{\delta m_0^2}{2p}(\sin 2\theta_0 
{\mathbf {\hat x}} - \cos 2 \theta_0 {\mathbf {\hat z}})\,,
\end{equation}
where $\delta m_0^2 = m_2^2 - m_1^2$ and $\theta_0$ are
the vacuum oscillation parameters.

The thermal potential from finite-temperature modification of the neutrino mass
due to the presence of thermally populated charged leptons in the plasma is
\begin{equation}
V^T(p) = -
\frac{8\sqrt{2} G_{\rm F} p}{3 m_{\rm W}^2} \left(\langle E_{l^-}
\rangle n_{l^-} + \langle E_{l^+} \rangle
n_{l^+}\right)\,.
\label{thermpot}
\end{equation}
The neutrinos also contribute a {\em thermal} self-potential similar to the
form of the self-potential on the right-hand side (RHS) of Eqs.~(\ref{veceqns})
\cite{Sigl}, but unlike $\alpha \sim G_F$, the thermal self-potential goes as
$G_F^2$.  Unless the initial asymmetry is of a size much too small to be
interesting here, the $G_F^2$ term is negligible by comparison with the order
$G_F$ self-potential and thus unimportant in determining the dynamics of the
system.

The background potential arising due to asymmetries in charged leptons
is nonzero only for electron neutrinos to maintain charge
neutrality of the baryon-contaminated plasma:
\begin{equation}
V^B =
\begin{cases}
\pm\sqrt{2}G_F\left(n_{e^-}-n_{e^+}\right) & \text{for\ }
\nu_e\rightleftharpoons\nu_{\mu,\tau},
\\
0 & \text{for\ }
\nu_\mu\rightleftharpoons\nu_\tau,
\end{cases}
\end{equation}
where $+(-)$ is for neutrinos (antineutrinos).  Due to the smallness
of the baryon asymmetry relative to number densities of thermalized
species, this term is always negligibly small relative to the vacuum
vector ${\mathbf \Delta}_p$ and the thermal potential $V^T$, and so we
may take ${\mathbf{\overline A}}_p = {\mathbf A}_p$.

In the absence of the neutrino self-potential it is possible to define
\begin{equation}
{\mathbf A}_p \equiv \frac{\delta m_m^2}{2p}(\sin 2\theta_m {\mathbf
  {\hat x}} - \cos 2 \theta_m {\mathbf {\hat z}}),
\end{equation}
where $\delta m_m^2$ and $\theta_m$ are the matter-affected
oscillation parameters.  When the self-term must be included, the
nonlinearity of the problem makes the notion of a matter-affected
mixing angle more subtle.


\subsection{Synchronization}

\begin{figure}
\begin{center}
\epsfig{file=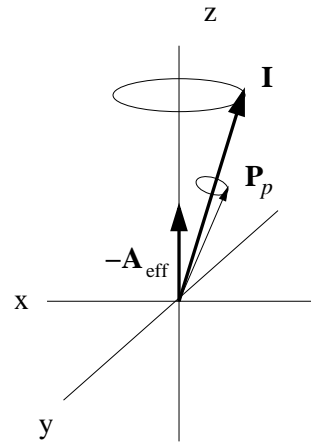,width=4cm} 
\caption{\label{vecpic} Vector precession diagram. The angles and
magnitudes of the vectors are not to scale but have been exaggerated
for clarity.  For the situation of interest, the magnitude of
${\mathbf I}$ is much greater than that of ${\mathbf A}_{\rm eff}$.
When this condition holds, it is a good approximation to describe the
evolution of the polarization vectors for the individual momentum
modes ${\mathbf P}_p$ as a precessing about ${\mathbf I}$.  The vector
${\mathbf I}$ then precesses about ${\mathbf A}_{\rm eff}$, in the
manner of a single momentum mode in the absence of the self-term.  For
asymmetries between neutrino flavors in the early universe, ${\mathbf
A}_{\rm eff}$ and ${\mathbf I}$ are both initially aligned with the
$z$ axis, and, for maximal mixing, both adiabatically evolve to align
with the $x$ axis.}
\end{center}
\end{figure}

Taking the difference of the Eqs.\ (\ref{veceqns}), integrating over
momenta, and defining a collective polarization vector ${\mathbf {I
\equiv J - \overline{J}}}$ one obtains
\begin{equation}
\partial_t{\mathbf I} \simeq {\mathbf A}_{\rm eff} \times {\mathbf I}, 
\label{iprecess}
\end{equation}
where the appropriate effective ``magnetic field'' is
\begin{equation}
\label{effA}
{\mathbf A}_{\rm eff} \simeq \frac{1}{{\mathbf I}^2} \int {\mathbf
A}_p ({\mathbf P}_p + {\mathbf{\overline P}}_p) \cdot{\mathbf I}.
\end{equation}
In fact, Eqs.~(\ref{iprecess}), (\ref{effA}) are exact only when
${\mathbf I}\parallel ({\mathbf P}_p + {\mathbf{\overline P}}_p)$. The
vector ${\mathbf I}$ thus precesses slowly about ${\mathbf A}_{\rm
eff}$.  Since the self-potential dominates Eqs.~(\ref{veceqns}), the
individual polarization vectors ${\mathbf P}_p$ all precess rapidly
about ${\mathbf I}$, and, if initially aligned, are held together.
The various vectors are illustrated in Fig.\ \ref{vecpic}.  

We assume for simplicity that the initial asymmetry resides in a
single flavor.  Although the coefficient of the self-term makes it
dominant, the flavor transformation is actually determined by the
evolution of ${\mathbf A}_{\rm eff}$.  In fact, if one leaves out the
self-term altogether, the synchronization is of course lost, but the
average flavor evolution of the system is almost completely unchanged.

Let us first consider the simple linearized case where the
self-potentials in Eqs.\ (\ref{veceqns}) vanish.  In this case, the
polarization vector of each momentum mode ${\mathbf P}_p$ precesses
about its own respective ${\mathbf A}_p$:
\begin{eqnarray}
\label{veceqns2}
\partial_t{\mathbf P}_p &=& +{\mathbf A}_p \times {\mathbf P}_p\,, \\
\nonumber \partial_t{\mathbf{\overline P}}_p &=& -{\mathbf{A}}_p
\times {\mathbf{\overline P}}_p\,.
\end{eqnarray}
If one follows only the average momentum $\langle p/T\rangle \simeq
3.15$, Eqs.~(\ref{veceqns2}) are simply two linear equations with a
straightforward solution.  Recall that each ${\mathbf P}_p$ is
initially aligned along the $+z$ axis. At high temperatures, $|V^T|\gg
|\mathbf \Delta_p|$, and thus from Eq.\ (\ref{magnetic}) the ${\mathbf
A}_p$ point in the $-\mathbf{\hat z}$ direction.  As the temperature
of the universe decreases, $|V^T| \sim T^5$ decreases and the vectors
${\mathbf A}_p$ will slowly rotate from the $-{\mathbf {\hat z}}$
direction toward the $+{\mathbf {\hat x}}$ direction and the angle
that ${\mathbf A}_p$ subtends with the $z$ axis will asymptote to
$2\theta_0$. This effect is a straightforward MSW transformation of
the asymmetry: the initial neutrino number excess in one flavor
evolves from a mass eigenstate in matter (modified by the thermal
potential) to a vacuum mass eigenstate with different flavor content.
And, since Eqs.\ (\ref{veceqns2}) are decoupled, the average-momentum
mode will describe the collective evolution of the entire system.

In the substantially more involved system, including the self-potential
(\ref{veceqns}), each momentum mode is coupled to all other momenta
through the self-term.  Therefore, a simplifying average-momentum
technique is poorly justified.  However, if one blindly drives forward
with an average-momentum evolution of Eqs.\ (\ref{veceqns}), one
luckily recovers (nearly) the correct behavior.  For the full-momentum
case, including the self-potential, the ${\mathbf I}$ vector will also
initially be aligned with the $z$ axis.  Using the approximate Eqs.\
(\ref{iprecess}) and making the as-yet unjustified assumption that
${\mathbf A}_{\rm eff}$ follows the {\it average} of the ${\mathbf
A}_p$, the synchronized system will undergo the MSW transformation at
the exact same temperature as the over simplified case
(\ref{veceqns2}).  This is what we found numerically.

The fact that the neutrino flavor system evolves to the same end-state
with and without the self-term appears at first absurd.  However, one
must keep in mind that the neutrino self-potential in a purely active
neutrino system affects the evolution drastically differently than the
familiar flavor-diagonal potentials in the matter Hamiltonian present,
for example, in evaluating the solar MSW effect and active-sterile
mixing in the early universe.  The neutrino self-potential, as the
dominant precession term for each momentum, only plays the role of
forcing each momentum mode to follow the collective vector
$\mathbf{I}$.

To explore the behavior of the system and verify the approximations in
arriving at the collective equations (\ref{iprecess}), (\ref{effA}), we
numerically integrate Eqs.\ (\ref{veceqns}), which, again, explicitly
couple the full thermal distribution of momenta to the quantum
mechanical evolution of each momentum mode.  Because of the
drastically different time scales over which the terms ${\mathbf
\Delta}_p$, $V^T(p)$ and $\alpha ({\mathbf J}- {\mathbf{\overline
J}})$ evolve, the system is a set of stiff nonlinear differential
equations and therefore requires careful treatment.

\begin{figure}
\begin{center}
\epsfig{file=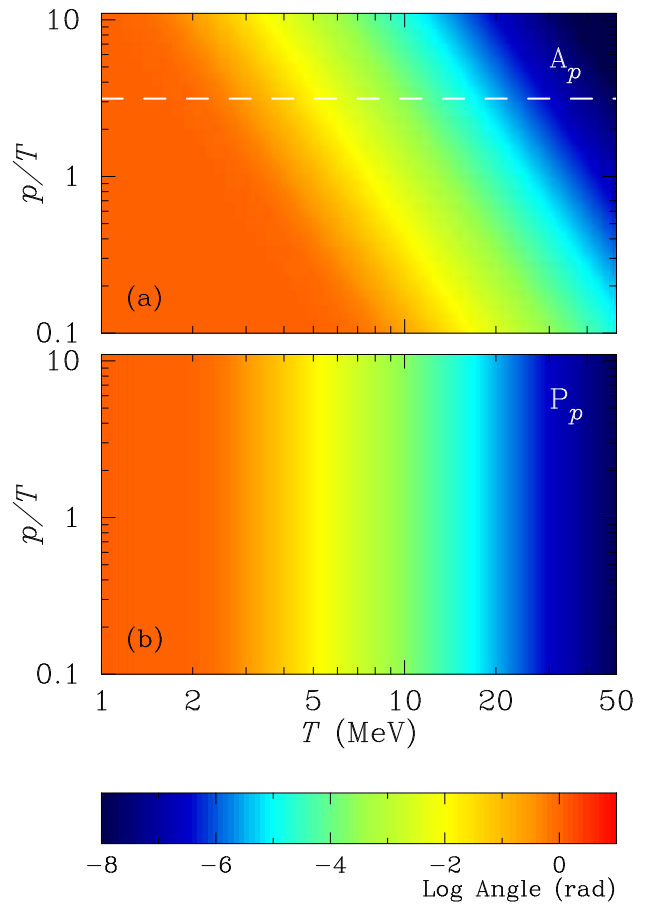,width=3.25in} 
\caption{\label{angpaper} The angle between ${\mathbf A}_p$ and the
$z$ axis is shown in the upper panel, in color, as a
function of the temperature of the universe (horizontally) and across
the neutrino spectrum (vertically).  In the lower panel, the angle
between ${\mathbf P}_p$ and the $z$ axis is displayed in the same
fashion.  As detailed in the text, all ${\mathbf P}_p$ ignore the
momentum dependence of ${\mathbf A}_p$ and are dramatically
synchronized to a single effective momentum, $p_{sync}/T \simeq \pi$.
That is, {\it all} of the ${\mathbf P}_p$ follow the orientation
(i.e., have the same color) of ${\mathbf A}_p$ at $p/T
\simeq \pi$, shown with a white horizontal dashed line.}
\end{center}
\end{figure}

\begin{figure}
\begin{center}
\epsfig{file=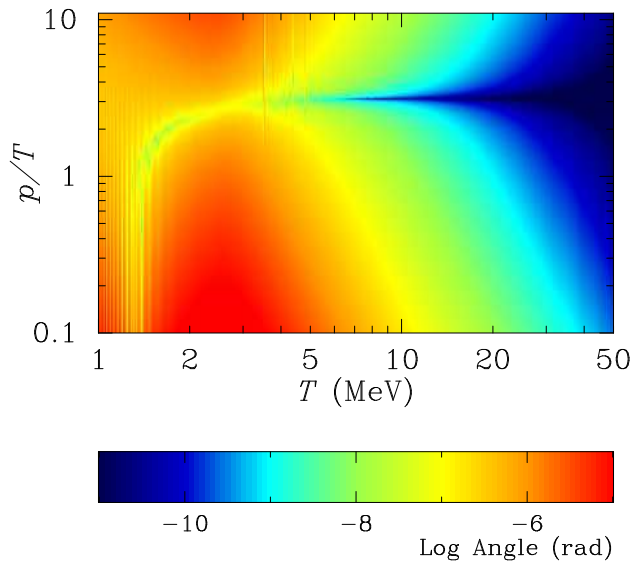,width=3.25in}
\caption{\label{angpi} We show the angle between individual ${\mathbf
P}_p$ and ${\mathbf I}$ as a function of temperature of the universe
(horizontally) and the neutrino spectrum (vertically).  The angles are
extremely small, indicating the degree of synchronization.}
\end{center}
\end{figure}

In Fig.\ \ref{angpaper}a we show the evolution with the full
equations~(\ref{veceqns}) of a representative two-flavor $\nu_e$ and $\nu_\mu$
system with the best-fit solar large mixing angle (LMA) parameters.  Shown is
the angle between each ${\mathbf A}_p$ and ${\mathbf {\hat z}}$, which would
determine evolution of each ${\mathbf P}_p$ in the absence of the self-term.
The actual angle between ${\mathbf P}_p$ and ${\mathbf {\hat z}}$ is shown in
Fig.~\ref{angpaper}b, displaying the stunning synchronization of all momentum
modes, to roughly the orientation of ${\mathbf A}_p$ at the average momentum
$\langle p/T \rangle \simeq 3$.  Figure~\ref{angpi} shows the tiny magnitude of
the angle between ${\mathbf P}_p$ and ${\mathbf I}$, which is the result of the
synchronization.

Now, we justify why taking the evolution of ${\mathbf A}_{\rm eff}$ to
be effectively that of the average of ${\mathbf A}_p$ luckily provides
the nearly the correct evolution.  For the case where collisional
damping may be neglected and assuming that both synchronization and
the close alignment of ${\mathbf P}_p+{\mathbf{\overline P}}_p$ with
${\mathbf I}$ holds, we may explicitly calculate ${\mathbf A}_{\rm
eff}$ which describes the MSW-like transition (the expressions for
${\mathbf A}_{\rm eff}$ and $p_{\rm sync}$ were independently
calculated in Ref.~\cite{yyy}).  We find
\begin{equation}
\label{int}
{\mathbf A}_{\rm eff} \simeq \xi \frac{\delta m_0^2}{2T} \left(
\frac{3/2}{\pi^2+\xi^2} \right) \Big[ \sin 2 \theta_0 {\mathbf{\hat
x}} + \left(-\cos 2 \theta_0 + Z \right) {\mathbf{\hat z}} \Big],
\end{equation}
where
\begin{equation}
Z= \frac{2T}{\delta m_0^2} \left(\frac{V^T}{p/T}\right)
\left( \pi^2 + \frac{\xi^2}{2} \right).
\end{equation}
Note that $Z$ is negative so there is no resonance.  It is helpful to
reexpress Eq.~(\ref{int}) in terms of effective ``synchronized''
oscillation parameters as
\begin{equation}
\label{Aint}
{\mathbf A}_{\rm{eff}} \equiv
\Delta_{\rm{sync}} \left( \sin 2 \theta_{\rm{sync}}  {\mathbf{\hat x}}
- \cos 2 \theta_{\rm{sync}} {\mathbf{\hat z}} \right),
\end{equation}
where the synchronized oscillation frequency is given by
\begin{equation}
\Delta_{\rm sync} = \xi \frac{\delta m_0^2}{2T} \left(
\frac{3/2}{\pi^2+\xi^2} \right)
\sqrt{\sin^2 2 \theta_0 + (-\cos 2 \theta_0 + Z)^2}.
\end{equation}
The size of $\Delta_{\rm{sync}}$, which is proportional to an overall
factor of $\xi$, is not important in determining when the MSW-like
transformation takes place.\footnote{If $V^T=0$, we find
$\Delta_{\rm{sync}} \simeq \delta m_0^2/132T$ for $\xi = 0.05$ in
agreement with Ref.~\cite{dhpprs}.  We note that the momentum scale
$p/T\simeq 132$ does not determine the character of the solution.}  It
is the mixing angle $\theta_{\rm sync}$ that is important in
describing the transformation resulting from the evolution into vacuum
mass eigenstates.  This angle (i.e., half of the angle between
${\mathbf A}_{\rm{eff}}$ with the $z$ axis) is given by
\begin{equation}
\sin^2 2 \theta_{\rm{sync}}
= \frac{\sin^2 2 \theta_0}{\sin^2 2 \theta_0 + (-\cos 2 \theta_0 +Z)^2},
\end{equation}
and thus we find it is the mixing angle which would correspond to
the momentum state
\begin{equation}
\frac{p_{\rm sync}}{T} = \pi \sqrt{1+\xi^2/2\pi^2}\simeq \pi.
\label{pi}
\end{equation}
This is one of our principal results, and indicates a remarkable
coincidence. Namely, the apparently identical evolution for the
synchronized system including the self-term and that found with a
vanishing self-term only results from the fact that average momentum
for a relativistic Fermi gas (with small chemical potential)
\begin{equation}
\langle p/T \rangle = \frac{7 \pi^4}{180 \zeta(3)} \simeq 3.15
\end{equation}
is approximately $\pi$, the effective momentum of the synchronized
system (\ref{pi}) with $\xi \ll \pi$.  One can observe in Fig.\
\ref{angpaper} that the effective mixing angle (the angle of ${\mathbf
A}_p$ with respect to the $z$ axis) does in fact correspond to the way
in which the state $p/T \simeq 3$ would evolve in the absence of the
self-term.

It is clear that $\theta_{\rm{sync}}$ depends only very weakly on the
size of the initial asymmetry, and in particular the transformation
will occur at almost the same temperature for any plausible initial
$\xi$.  Additionally, if $\xi$ were very large, even a very small
degree of flavor transformation would be sufficient to upset
successful BBN.  We have also numerically integrated the system for
several initial asymmetries, and verified that the synchronized
transformation is present for all asymmetries within the previous
limit in Eq.\ (\ref{oldlimits}).

Note that a large mixing angle is essential in obtaining flavor
``equilibration,'' i.e., that $\xi_\mu^i$ is effectively transferred
to $\xi_e^f$ as shown by DHPPRS~\cite{dhpprs}.  The underlying
dynamics is simply the adiabatic evolution of the initial neutrinos
into vacuum mass eigenstates.  This would be exactly the usual MSW
effect\footnote{Note, however, that for a normal hierarchy we do not
have a resonance --- the negative thermal potential makes the $\nu_e$'s
{\it lighter}, and does the same for $\bar\nu_e$.} if it were not for
additional complexity of synchronization.  We achieve equilibration in
the sense that the initial asymmetry is partitioned across the flavors
(with the ratio of the final $\xi_e^f$ and $\xi_{\mu,\tau}^f$ set by the
vacuum mixing angle).  This ``equilibration'' is simply a MSW
transformation that leaves the ensemble in a coherent state.  This is
to be distinguished from equilibration in the conventional sense of a
completely incoherent or relaxed state, i.e., one produced by
collisions.


\section{Neutrino Properties and Asymmetry Transformation}

Since the asymmetry in the electron neutrino number is the most
stringently constrained, its enhancement due to coupling to the other
flavors is crucial.  The neutrino oscillation solution best fitting
the observed solar electron neutrino flux and spectra is the region of
mixing parameter space named the large mixing angle (LMA) solution,
with maximum likelihood parameters for two-neutrino mixing of order
$\delta m_0^2 \approx 4\times 10^{-5}\rm\ eV^2$ and $\sin^2
2\theta_0\approx 0.8$.  Since the LMA mixing is large but not maximal,
the first mass state ($|m_1\rangle$) is more closely associated with
the electron neutrino and the other ($|m_2\rangle$) less so, and in
order to enable resonance in the sun, $m_1<m_2$.

We also know from atmospheric neutrino observations that $\mu$ and
$\tau$ neutrino flavors are maximally mixed superpositions (or nearly
so) of two mass states $|m_2\rangle$ and $|m_3\rangle$.  Therefore,
the flavor composition of mass state $|m_2\rangle$ is that of a nearly
maximal superposition, and complicates discussion of LMA mixing in
neutrino environments where the flavor content of $|m_2\rangle$ is of
interest.

However, a powerful simplification can be made given maximally mixed
(or more generally ``similarly coupled'') $\nu_\mu$ and $\nu_\tau$,
which allows a linear transformation of the $3\times 3$ mixing matrix
such that one effective flavor state $|\nu_\tau^\ast\rangle$ is
identically a vacuum mass eigenstate and decouples from the matter
effects~\cite{transform}.  It is sufficient to follow the two
remaining states $|\nu_e\rangle$ and $|\nu^\ast_\mu\rangle$.  This is
only justified if the momentum state (or more exactly, momentum
distributions) of the superimposed flavors $|\nu_\mu\rangle$ and
$|\nu_\tau\rangle$ are indistinguishable.  That is, the temperatures
and chemical potentials of $\nu_\mu$ and $\nu_\tau$ should be equal,
i.e., $\nu_\mu$, $\nu_\tau$ should be equilibrated.  The atmospheric
neutrino results actually can provide that $\nu_\mu$ and $\nu_\tau$
are equilibrated, which we discuss in Section \ref{atmsec}.  Strictly
speaking, the similar-coupling limit is exact only when $U_{e3} = 0$,
and we consider the more general case below.


\subsection{Electron flavor transformation}

Recall that we are interested in whether an asymmetry initially
present in the poorly constrained $\nu_\mu$ or $\nu_\tau$ will convert
into a stringently constrained $\nu_e/\bar\nu_e$ asymmetry.  We can
analyze how the LMA mixing parameters evolve in the early universe
through the effective two-neutrino system $(\nu_e,\nu_\mu^\ast)$.  The
initial system may be prepared by an unspecified leptogenesis
mechanism to be in an unmixed state with the asymmetry in
$\nu_\mu^\ast$ ($\xi_{\mu^\ast}^i \neq 0$) and no asymmetry in $\nu_e$
($\xi_e^i = 0$) and remains in this state from damping by collisions.
We have defined the direction $+\mathbf{\hat z}$ to correspond to the
$\nu_e$ flavor.  Taking, for the sake of the example, the initial
$\xi_{\mu^\ast}^i$ to be negative, the vector ${\mathbf I}$ will
initially point in the $+\mathbf{\hat z}$ direction.

At initially high temperatures, the effective magnetic field vector ${\mathbf
A}_{\rm eff}$ is dominated by the thermal lepton potential $V^T$, and is
aligned in the $-\mathbf{\hat z}$ direction.  As the universe cools, ${\mathbf
A}_{\rm eff}$ rotates away from $-\mathbf{\hat z}$ and asymptotes to its vacuum
value, which lies close to the $+\mathbf{\hat x}$ direction (i.e., the angle it
makes with the $z$ axis is $2\theta_0$).  A large vacuum mixing angle is
clearly necessary for this MSW transformation to work.

\begin{figure}
\begin{center}
\epsfig{file=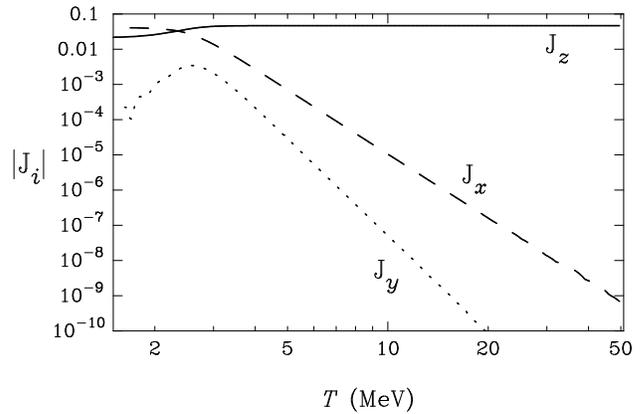,width=3.25in} 
\caption{\label{pilma} The evolution of $J_i$ in the synchronized case
with LMA parameters.  As described in the text, the behavior is
essentially a MSW transformation
$\nu_e\leftrightarrow\nu^\ast_\mu$. The antineutrinos $|\bar J_i|$
evolve identically.  {\it The fact that $J_y$ is never large
demonstrates that all of the precession angles are small enough that
the evolution is dominantly in the $x-z$ plane}.  The evolution of
$P_i$ and $\bar P_i$ at the average momentum is the same if one
excludes the self-potentials in Eq.~(\ref{veceqns}).}
\end{center}
\end{figure}

For an initial asymmetry of $|\xi_{\mu^\ast}^i| = 0.05$, the evolution of
the synchronized vector components $J_i$ are shown in Fig.\
\ref{pilma}.  The components are driven as a magnetic dipole
adiabatically following the evolution of the magnetic field ${\mathbf
A}_{\rm eff}$.  {\it The evolution of $P_i (\bar P_i)$ at the average
momentum is the same if one excludes the self-potentials in
Eq.~(\ref{veceqns}).}  The power law growth of $J_x$ is simply the
evolution of the synchronized mixing angle
\begin{equation}
\label{powerlaw}
J_x \sim 2 \theta_{\rm{sync}}\sim \frac{1}{Z}
\equiv \frac{\delta m_0^2/2p_{\rm sync}}{V^T(p_{\rm sync})}
\sim T^{-6}\,,
\end{equation}
and the growth of $J_y \sim T^{-9}$ results directly from
(\ref{iprecess}).  Obviously, the transformation occurs when the
orientation of ${\mathbf J}$ rapidly evolves at $\langle
\Delta_p\rangle \sim \langle V^T\rangle$ at $T\sim 2\rm\ MeV$.  The
temperature of the transition point in $J_x$ scales only as $(\delta
m^2)^{1/6}$, so the results are rather insensitive to the uncertainty
in the LMA $\delta m^2$.  The antineutrinos evolve identically by
following the vector $-{\mathbf A}_{\rm eff}$.  The final state of the
asymmetry after the MSW transformation entering the nucleosynthesis
epoch is then transferred in proportion to the vacuum mixing amplitude
between the two flavors, i.e., $J_z(1\rm\ MeV)$:
\begin{eqnarray}
\xi_e^f &=& 
\left(\frac{1-\cos 2\theta_0}{2}\right)\xi_{\mu^\ast}^i \,, \\
\xi_{\mu^\ast}^f &=& 
\left(\frac{1+\cos 2\theta_0}{2}\right)\xi_{\mu^\ast}^i \,.
\label{xief}
\end{eqnarray}
(Some care must be taken in interpreting the limits on $\xi^f$, since the final
distributions are not exactly thermal, as they are superpositions of
Fermi-Dirac distributions with different chemical potentials.) Obviously,
complete ``equilibration'' or $\xi_e^f=\xi_{\mu^\ast}^f$ only occurs for
maximal mixing.  The antineutrino chemical potential evolves to the values
$\bar\xi^f_e=-\xi_e^f$ and $\bar\xi^f_\mu=-\xi_\mu^f$.  We note that collisions
(which we have neglected) will help make the flavor transformation more
complete and thus should reduce the sensitivity to the mixing angle.

\subsection{Mu-tau flavor transformation}\label{atmsec}

\begin{figure}
\begin{center}
\epsfig{file=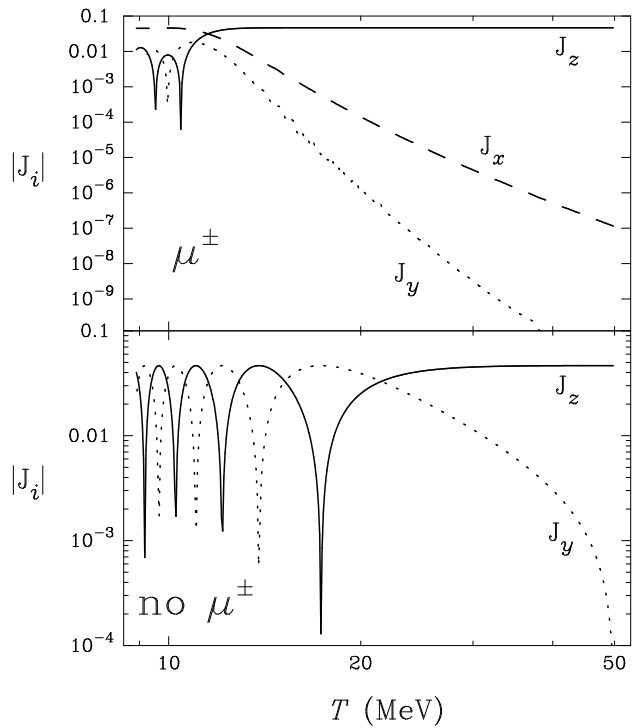,width=3.25in} 
\caption{\label{piatm} The evolution of $J_i$ ($\bar J_i$ are
identical) for the mu and tau neutrino transformation with and without the
inclusion of thermal $\mu^\pm$ pairs.  The spiky features indicate
real oscillations going through zero, and the depth of the spikes on
the logarithmic scale is an artifact of numerical sampling.  Those
oscillations are real and are determined by the atmospheric $\delta
m^2_0$.  In the lower panel, $J_x$ is zero since the mixing angle is
maximal. Collisions have been ignored.}
\end{center}
\end{figure}

Maximal neutrino mixing as indicated by the Super-Kamiokande observations of
atmospheric neutrinos has nearly identical implications for the evolution of
asymmetries between $\nu_\mu$ and $\nu_\tau$.  Because of the hierarchy $\delta
m^2_{\rm atm}\gg \delta m^2_{\rm LMA}$, $\langle\Delta_p\rangle\sim \langle
V^T\rangle$ at the higher temperature $T\sim 10\rm\ MeV$. Necessary in driving
the flavor evolution here is the presence of the remnant thermally produced
charged muons with energy density
\begin{equation}
\rho_{\mu^\pm} =
\frac{1}{\pi^2} \int{p^2 dp\;\frac{\sqrt{p^2+m_\mu^2}}
{1+\exp\left(\sqrt{p^2+m_\mu^2}/T\right)}}\,.
\end{equation}
Though far from the thermal abundance of $e^\pm$ at $T\sim 20\rm\
MeV$, real muons remain enough to dictate the flavor evolution.
However, since $T < m_\mu$, the thermal potential from Eq.\
(\ref{thermpot}) is modified as $\langle E_{\mu^\pm} \rangle
\rightarrow \frac{3}{4}\langle  E_{\mu^\pm} + p^2/3
E_{\mu^\pm}\rangle$ \cite{nrdnt}.  We solved the evolution of this
case numerically, explicitly including the thermal abundance of
$\mu^\pm$, whose disappearance accelerates the growth of $J_x$ and
$J_y$ away from the power-law growth in the previous LMA case, but for
simplicity have ignored collisions (Fig.\ \ref{piatm}).  Maximal
mixing then gives an equilibration $\xi_{\mu}^f = \xi_{\tau}^f$, which
allows the application of the simplifying basis transformation of the
previous section.  Inclusion of collisions would damp the oscillations
at low temperatures but not the transformation, as found by
DHPPRS~\cite{dhpprs}.

Interestingly, in the case of evolution without the presence of
thermal $\mu^\pm$, the evolution is different, with pure synchronized
vacuum oscillations taking place (rotation in the $z-y$ plane) after
the Hubble time exceeds the oscillation time.  Collisions, which we
have omitted, would modify the oscillations seen here.

\subsection{Effects of $U_{e3}$}

The possibility of a nonzero value of $U_{e3}$ obstructs the
simplifying linear transformation to the basis $|\nu_e\rangle$ and
$|\nu^\ast_\mu\rangle$.  However, nonzero $U_{e3}$ may allow partial
equilibration of $\xi_\mu,\xi_\tau$ into $\xi_e$ earlier, at $T \sim 5
{\rm\ MeV}$.  For solar LMA mixing, significant transformation will
always occur at $T \sim 2 {\rm\ MeV}$ so the value of $U_{e3}$ will
not alter the basic outcome.  However, substantial equilibration at 5
MeV (well before the beta-equilibrium freeze-out) makes the general
conclusions even more inevitable.

\begin{figure}
\begin{center}
\epsfig{file=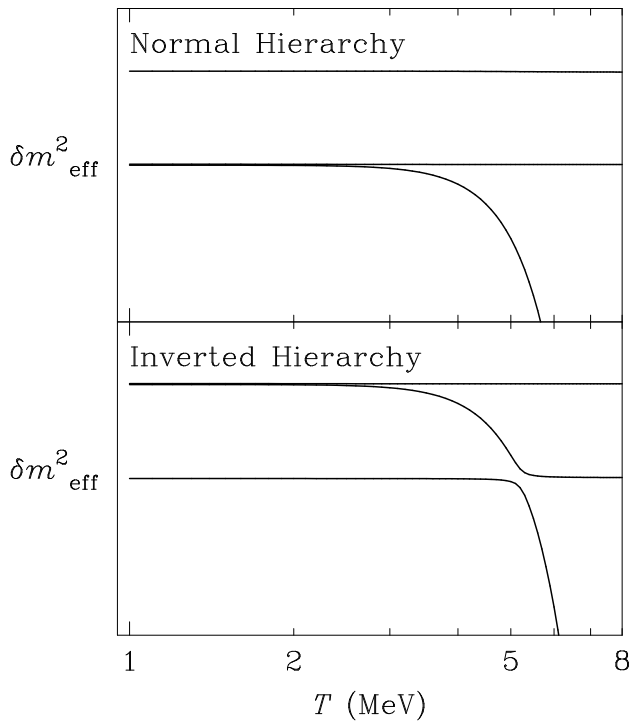,width=3.25in} 
\caption{\label{level} Level-crossing diagrams for neutrinos of the average
momentum in the absence of the self-potential.  In the upper panel we have a
normal hierarchy, where the neutrino mass eigenstates asymptote to their vacuum
values, without ever going through a resonance.  This is to be contrasted with
the inverted hierarchy shown in the lower panel where nonzero $U_{e3}$ leads to
a resonance at $T\sim 5 {\rm\ MeV}$.  }
\end{center}
\end{figure}

There are, however, some subtleties associated with the sign of
$\delta m^2_{\rm atm}$ --- that is, whether the neutrino spectrum has a
normal or inverted hierarchy.\footnote{The sign of the solar $\delta
m^2$ is determined by the requirement that there be a MSW transition in
the Sun, which precludes a resonance in the early universe (for both
neutrinos and antineutrinos).  There is, however, no such
constraint of the sign of the atmospheric $\delta m^2$.}  For a normal
hierarchy, the fact the thermal potential makes the $\nu_e$'s (and
$\overline{\nu}_e$'s) lighter implies that that no resonance
conditions can be satisfied in the early universe. With an inverted
hierarchy, however, a $\nu_e - \nu^\ast_\mu$ resonance will occur when
$V^T\sim \delta m^2_{\rm atm}$.  We plot in Fig.\ \ref{level}
level crossing diagrams for neutrinos of the average energy in the
absence of the self potential.  As discussed above, this is a very
good description of the evolution of the entire neutrino distribution.

The $U_{e3}$ mixing angle is constrained to be small~\cite{choozpv},
and as such, coherent evolution will not lead to large flavor
transformation (for the inverted hierarchy, coherent evolution through
the resonance would swap asymmetries between flavors).  However, at $T
\sim 5 {\rm\ MeV}$ collisional processes are still highly important
and will help achieve equilibration.  This should be somewhat more
effective for the inverted case (where the mixing angle goes through a
maximum) as collisions equilibrate most effectively when mixing angles
are large.

\section{New Constraints}

The electron neutrino asymmetry $\xi_e$ is limited by its effects on
the primordial $^4$He abundance, $Y_p$.  At nucleosynthesis, nearly
all neutrons are incorporated into $^4$He nuclei, and $Y_p$ production
is limited by the neutron fraction, set by the freeze-out of
beta equilibrium at $T \simeq 1\rm\ MeV$.  The change in the neutron
to proton ratio with non-zero $\xi_e$ is simply a Boltzmann factor
$n/p\propto e^{-\xi_e} \approx 1-\xi_e$.  And since $Y_p\propto n/p$,
the uncertainty in the constraint on $\xi_e$ is directly related to
the uncertainty in the primordial helium abundance $\Delta Y_p$,
\begin{equation}
\Delta \xi_e \approx \frac{\Delta Y_p}{Y_p}.
\end{equation}
Therefore, one can be very conservative regarding the error on the
primordial abundance, e.g., $\Delta Y_p\approx \pm 0.010$ \cite{os}
and still limit $\Delta \xi_e \approx \pm 0.04$, or equivalently,
$|L_e| \lesssim 0.03$.  This method ultimately relies on the
uncertainty (mostly systematic) in the primordial abundance of $^4$He.
Refinement of this constraint may be possible by applying CMB priors
to BBN predictions combined with reduced systematic uncertainties of
observed primordial element abundances \cite{Cyburt}.

Analysis of synchronized transformation of neutrino asymmetries
indirectly translates the constraints on $\xi_e^f$ to $\xi_\mu^i$ and
$\xi_\tau^i$.  In the extreme scenario, an asymmetry
$\xi_{\mu,\tau}^i$ in $\nu_\mu$ ($\nu_\tau$) is equilibrated with
$\nu_\tau$ ($\nu_\mu$) for maximal mixing, such that the state
$\nu^\ast_\mu$ has $\xi_{\mu^\ast} = 0.5 \xi_{\mu,\tau}^i$.  The LMA
solution transforms $\xi_{\mu^\ast}$ as Eq.~(\ref{xief}) so that
\begin{equation}
\xi^f_e = \left(\frac{1-\cos 2\theta_0}{4}\right)\xi_{\mu,\tau}^i.
\end{equation}
For the best-fit LMA mixing angle $\sin^2 2\theta_0\approx 0.8$, the limit on
an initial asymmetry is $\xi_{\mu,\tau}^i \lesssim 0.3$.  However, the LMA
mixing angle is not precisely specified.  The lower end of the 95\% confidence
level (C.L.) region has $\sin^2 2\theta_0\approx 0.6$, for which the limit on
the initial asymmetry is considerably weaker\footnote{Note that we expect this
limit would be tighter were we to include the effect of collisions.}
$\xi_{\mu,\tau}^i \lesssim 0.5$.  The effective ``$2\sigma$'' limit therefore
is actually an order of magnitude larger than that given in DHPPRS since a
``small-angle'' LMA solution reduces the transformation amplitude considerably.
The sensitivity of the KamLAND experiment to the LMA parameter space can
confirm the LMA parameters~\cite{Piepke} and potentially reduce the mixing
angle uncertainty, and thus improve constraints on the lepton number.


\section{Discussion and Conclusions}

Due to synchronization by the neutrino self-potential, transformation
of a large fraction of any asymmetries in $\nu_\mu$ or $\nu_\tau$
number to $\nu_e$ is an inescapable consequence of the near
bimaximal mixing framework emerging for the neutrino mass matrix.  We
have performed a full numerical integration of the evolution equations
in Eq.\ (\ref{veceqns}).  The numerical solution is nontrivial due to
stiff, nonlinear equations with terms whose time scales vary by
several orders of magnitude.  We confirm the numerical results of
DHPPRS, and agree that large initial asymmetries in $\nu_\mu$ and
$\nu_\tau$ are effectively transformed into a $\nu_e$ asymmetry, so
that the bound from BBN bounds all \cite{dhpprs}.  

In addition, we have shown numerically that the coupled evolution of
the full-momentum results can also be obtained in the average-momentum
case when the nonlinear coupling is neglected.  The transfer of
neutrino asymmetries between flavors occurs identically even when
ignoring the numerically dominant self-potential.  In Eq.\ (\ref{pi})
we have derived that the self-potential drives a synchronization of
all momenta to a momentum mode $p/T = \pi$, so that the system by
numerical coincidence closely follows the average momentum case $p/T
\simeq 3.15$. 

We conclude by considering the following implications of these results:

(1) The uncertainty in the lepton number of the universe may be
reduced by up to two orders of magnitude.  However, the most
conservative limits place 
\begin{eqnarray}
\label{newlimits}
|\xi_e^f| &\lesssim& 0.04\,, \\
|\xi_\mu^i+\xi_\tau^i| &\lesssim& 0.5\,,
\end{eqnarray}
($|L_e| \lesssim 0.03$ and $|L_\mu+L_\tau| < 0.4$). These limits will be
improved by reducing systematic uncertainties in the inferred primordial $^4$He
abundance and the precise determination of the baryon density by satellite
anisotropy experiments Microwave Anisotropy Probe (MAP) and Planck
\cite{hudodelson}.  It also may be improved by verification of the LMA
parameters by KamLAND, particularly if the mixing angle is at the large end of
the presently allowed range.  The upcoming data from SNO~\cite{sno} will also
play a very important role in reducing the mixing parameter uncertainties.

(2) Because effectively asymmetries in any neutrino flavor will
affect beta equilibrium, the stringent limits (\ref{newlimits})
consequentially eliminate the possibility of degenerate BBN
\cite{oritoesposito}, since an increase the expansion rate with large
$|\xi_{\mu,\tau}|\sim 1$ can no longer be compensated by a small
$\xi_e\sim 0.1$.

(3) The above limits on degeneracy in terms of extra relativistic degrees of
freedom $\Delta N_\nu$ [see Eq.~(\ref{delnu})] are impressively small: $\Delta
N_{\nu}\lesssim 0.004$ for the best-fit LMA solution, and $\Delta
N_{\nu}\lesssim 0.2$ for the lower limit on the mixing angle in the LMA
solution.  DHPPRS suggest that $\Delta N_\nu$ can be eliminated as a
cosmological parameter in upcoming fits to the precision CMB
data~\cite{leptcmb}.  It is certainly true that $\xi$ can be eliminated, but
that is not the only possible contribution to $\Delta N_\nu$.  {\it If any
  nonstandard contribution to the relativistic energy density were to be
  detected via the CMB, its origin would be something more exotic than
  degenerate neutrinos}, e.g., the decay of a massive particle to relativistic
species after BBN but before CMB decoupling~\cite{Hannestad:2001hn}.

(4) It is actually still possible that the upper limit for $\xi_e$
in Eq.~(\ref{oldlimits}) be fulfilled.  Strictly speaking, we have set
tight new degeneracy limits assuming no non-standard contribution to
the energy density at the time of BBN.  It is conceivable that $\xi_e
\sim \xi_{\mu} \sim \xi_{\tau} \sim 0.2$ if another relativistic
particle or scalar field contributes the extra energy density required
to compensate for the large $\nu_e$ chemical potential.  In this case,
flavor-transformation improves the current $\xi_{\mu,\tau}$ limits by
at most an order of magnitude.  Such an unnatural scenario can be
detected by comparison with the CMB.

(5) A possible complication to the scenario presented here could be
mixing with a light sterile neutrino.  Obviously, if the LSND result
is confirmed by MiniBooNE, then the physics will be much more
complicated than assumed here.  If the LSND result is not confirmed,
there is still the possibility of subdominant mixing to steriles that
may be difficult to detect in neutrino oscillation experiments, but
which may still play an important role in the early universe.  Such
scenarios have not yet been explored.

(6) A final complication is the yet-unexcluded possibility of a low
reheating temperature ($T\sim 1\rm\ MeV$) \cite{lowheat}, such that
the initial conditions of thermal or chemical equilibrium for
neutrinos for the analysis presented here is invalid.  Stronger
constraints on low-temperature reheating scenarios may be obtained by
studying their effects on the light element abundances in detail.


\section*{Acknowledgments}

We thank Scott Dodelson, George Fuller, Ray Sawyer, Ray Volkas, and Yvonne Wong
for useful discussions.  K.N.A., J.F.B., and N.F.B. were supported by Fermilab,
which is operated by URA under DOE contract No.\ DE-AC02-76CH03000, and were
additionally supported by NASA under NAG5-10842.



\begin{thebibliography}{99}

\bibitem{schramm}
D.~N.~Schramm and M.~S.~Turner, Rev.\ Mod.\ Phys.\  {\bf 70}, 303 (1998).

\bibitem{cmb} 
C.~Pryke, N.~W.~Halverson, E.~M.~Leitch, J.~Kovac, J.~E.~Carlstrom,
W.~L.~Holzapfel, and M.~Dragovan, 
Astrophys.\ J.\  {\bf 568} (2002) 46;
P.~de Bernardis {\it et al.}, {\it ibid.} {\bf 564} (2002) 559.

\bibitem{Turner}
M.~S.~Turner, astro-ph/0106035.

\bibitem{lss}
W.~J.~Percival {\it et al.}, Mon.\ Not.\ R.\ Astron.\ Soc. {\bf 327},
1297 (2001).

\bibitem{affleckdine}
I.~Affleck and M.~Dine, Nucl.\ Phys.\ {\bf B249}, 361 (1985); \\
A.~Casas, W.~Y.~Cheng, and G.~Gelmini, {\it ibid.} 297 (1999).

\bibitem{ftv}
R.~Foot, M.~J.~Thomson, and R.~R.~Volkas, Phys.\ Rev.\ D {\bf 53}, 5349 (1996).

\bibitem{Hannestad:2000hc}
S.~Hannestad, Phys.\ Rev.\ Lett.\  {\bf 85}, 4203 (2000).

\bibitem{leptcmb}
W.~H.~Kinney and A.~Riotto, Phys.\ Rev.\ Lett.\  {\bf 83}, 3366 (1999); \\
R.~E.~Lopez, S.~Dodelson, A.~Heckler, and M.~S.~Turner,
{\it ibid.}  {\bf 82}, 3952 (1999).

\bibitem{hansen}
S.~H.~Hansen, G.~Mangano, A.~Melchiorri, G.~Miele, and O.~Pisanti,
Phys.\ Rev.\ D {\bf 65}, 023511 (2002).

\bibitem{kneller}
J.~P.~Kneller, R.~J.~Scherrer, G.~Steigman and T.~P.~Wal\-ker,
Phys.\ Rev.\ D {\bf 64}, 123506 (2001).

\bibitem{solar}
Homestake Collaboration, B.~T.~Cleveland {\it et al.},
Astrophys.\ J.\  {\bf 496}, 505 (1998);\\
GALLEX Collaboration, W.~Hampel {\it et al.},
Phys.\ Lett.\ B {\bf 447}, 127 (1999);\\
SAGE Collaboration,J.~N.~Abdurashitov {\it et al.}, 
Phys.\ Rev.\ C {\bf 60}, 055801 (1999);\\
Super-Kamiokande Collaboration, S.~Fukuda {\it et al.},
Phys.\ Rev.\ Lett.\  {\bf 86}, 5651 (2001);  5656 (2001);\\
GNO Collaboration, M.~Altmann {\it et al.}  ,
Phys.\ Lett.\ B {\bf 490}, 16 (2000);\\
SNO Collaboration, Q.~R.~Ahmad {\it et al.},
Phys.\ Rev.\ Lett.\  {\bf 87}, 071301 (2001).

\bibitem{atm}
Super-Kamiokande Collaboration, S.~Fukuda {\it et al.} ,
Phys.\ Rev.\ Lett.\  {\bf 85}, 3999 (2000);\\
Super-Kamiokande Collaboration, Y.~Fukuda {\it et al.},
{\it ibid.} {\bf 81}, 1562 (1998).

\bibitem{lsnd}
LSND Collaboration, A.~Aguilar {\it et al.},
Phys.\ Rev.\ D {\bf 64}, 112007 (2001).

\bibitem{fournu}
M.~C.~Gonzalez-Garcia and Y.~Nir, hep-ph/0202058; \\
C.~Giunti, J.\ High Energy Phys.\ {\bf 0001}, 032 (2000).

\bibitem{dibarietal}
P.~Di Bari, Phys.\ Rev.\ D {\bf 65}, 043509 (2002); \\
N.~F.~Bell, R.~Foot, and R.~R.~Volkas, 
{\it ibid.} {\bf 58}, 105010 (1998); \\
K.~Abazajian, G.~M.~Fuller, and X.~Shi,
{\it ibid.} {\bf 62}, 093003 (2000).

\bibitem{miniboone}
MiniBooNE Collaboration, A.~Bazarko,
Nucl.\ Phys.\ B (Proc.\ Suppl.)  {\bf 91}, 210 (2000).

\bibitem{smirnov}
C.~Lunardini and A.~Y.~Smirnov, Phys.\ Rev.\ D {\bf 64}, 073006 (2001).

\bibitem{smf}
M.~J.~Savage, R.~A.~Malaney and G.~M.~Fuller,
Astrophys.\ J.\  {\bf 368}, 1 (1991).

\bibitem{dhpprs} 
(DHPPRS): A.~D.~Dolgov, S.~H.~Hansen, S.~Pastor, S.~T.~Petcov,
G.~G.~Raffelt, and D.~V.~Semikoz, Nucl.\ Phys.\ {\bf B632}, 363 (2002).

\bibitem{oritoesposito}
H.~S.~Kang and G.~Steigman, Nucl.\ Phys.\ {\bf B372}, 494 (1992); \\
S.~Esposito, G.~Mangano, G.~Miele, and O.~Pisanti,
J.\ High Energy Phys.\ {\bf 0009}, 038 (2000); \\
M.~Orito, T.~Kajino, G.~J.~Mathews and Y.~Wang, Phys.\ Rev.\ D {\bf 65}, 123504
(2002). 

\bibitem{stodolsky}
L.~Stodolsky, Phys.\ Rev.\ D {\bf 36}, 2273 (1987); \\
B.~H.~McKellar and M.~J.~Thomson, {\it ibid.} {\bf 49}, 2710 (1994).

\bibitem{pantaleone}
J.~Pantaleone, Phys.\ Lett.\ B {\bf 287}, 128 (1992).

\bibitem{prs}
S.~Pastor, G.~G.~Raffelt, and D.~V.~Semikoz,
Phys.\ Rev.\ D {\bf 65}, 053011 (2002).

\bibitem{linearsync}
N.~F.~Bell, R.~F.~Sawyer, and R.~R.~Volkas,
Phys.\ Lett.\ B {\bf 500}, 16 (2001).

\bibitem{Samuel}
S.~Samuel, Phys.\ Rev.\ D {\bf 48}, 1462 (1993); \\
V.~A.~Kostelecky and S.~Samuel, {\it ibid.} {\bf 52}, 621 (1995); \\
S.~Samuel, {\it ibid.} {\bf 53}, 5382 (1996); \\
J.~Pantaleone, {\it ibid.} {\bf 58}, 073002 (1998).

\bibitem{Sigl}
G.~Sigl and G.~Raffelt, Nucl.\ Phys.\ B {\bf 406}, 423 (1993).

\bibitem{yyy}
Y.~Y.~Y.~Wong, Phys.\ Rev.\ D (to be published), hep-ph/0203180.

\bibitem{transform}
D.~O.~Caldwell, G.~M.~Fuller, and Y.~Z.~Qian,
Phys.\ Rev.\ D {\bf 61}, 123005 (2000); \\
A.~B.~Balantekin and G.~M.~Fuller,
Phys.\ Lett.\ B {\bf 471}, 195 (1999).

\bibitem{nrdnt}
D.~Notzold and G.~Raffelt,
Nucl.\ Phys.\ {\bf B307}, 924 (1988); \\
J.~C.~D'Olivo, J.~F.~Nieves and M.~Torres,
Phys.\ Rev.\ D {\bf 46}, 1172 (1992).

\bibitem{choozpv}
CHOOZ Collaboration, M.~Apollonio {\it et al.},
Phys.\ Lett.\ B {\bf 466}, 415 (1999); \\
Palo Verde Collaboration, F.~Boehm {\it et al.},
Phys.\ Rev.\ D {\bf 64}, 112001 (2001).

\bibitem{os}
K.~A.~Olive and E.~Skillman, New Astron. {\bf 6}, 246 (2001).

\bibitem{Cyburt}
R.~H.~Cyburt, B.~D.~Fields, and K.~A.~Olive,
Astropart.\ Phys.\  {\bf 17}, 87 (2002).

\bibitem{Piepke}
KamLAND Collaboration, A.~Piepke,
Nucl.\ Phys.\ B (Proc.\ Suppl.)  {\bf 91}, 99 (2001).

\bibitem{hudodelson}
W.~Hu and S.~Dodelson, astro-ph/0110414.

\bibitem{sno}
SNO Collaboration, J.~R.~Klein, hep-ex/0111040.

\bibitem{Hannestad:2001hn}
S.~Hannestad, Phys.\ Rev.\ D {\bf 64}, 083002 (2001).

\bibitem{lowheat}
M.~Kawasaki, K.~Kohri, and N.~Sugiyama,
Phys.\ Rev.\ Lett.\  {\bf 82}, 4168 (1999); \\
G.~F.~Giudice, E.~W.~Kolb, and A.~Riotto,
Phys.\ Rev.\ D {\bf 64}, 023508 (2001).

\end{thebibliography}
\end{document}